\documentclass[11pt]{article}

\usepackage{amsmath,amssymb,amsthm}
\usepackage[margin=1in]{geometry}
\usepackage{booktabs}
\usepackage{algorithm}
\usepackage{algpseudocode}
\usepackage[authoryear,round]{natbib}
\usepackage[colorlinks=true,citecolor=blue,linkcolor=blue,urlcolor=blue]{hyperref}
\usepackage{graphicx}
\usepackage{tabularx}
\usepackage{bbm}
\usepackage{bm}
\usepackage{tikz}
\usetikzlibrary{positioning,arrows.meta,backgrounds}

\newtheorem{proposition}{Proposition}

\newtheorem{theorem}{Theorem}
\newtheorem{corollary}{Corollary}

\begin{document}

\title{Competitive Balance Beyond Standings: Bayesian Inference for Latent Tiers and Structural Heterogeneity}
\author{Guanyu Hu\\
Department of Statistics and Probability\\
Michigan State University\\
\texttt{huguanyu@msu.edu}}
\date{}
\maketitle

\begin{abstract}
Competitive balance has long been a central concern in sports economics, tied to outcome uncertainty, fan engagement, and the way leagues design their policies. Most existing measures of competitive balance are built from standings or winning percentages, which say little about how overall parity relates to the way teams sort into latent competitive tiers. We propose a Bayesian framework based on the Bradley–Terry model that jointly estimates the number of competitive tiers and team memberships. To capture the overall degree of competitive heterogeneity, we introduce a structural balance parameter defined on the unit interval within a hierarchical model, and we establish theoretical results giving this parameter a clear interpretation: it is directly tied to the separation between tiers, and overall heterogeneity depends on both the size and frequency of cross-tier differences. Importantly, structural balance is distinct from the information content of the schedule itself, so leagues with different numbers of teams and games can be compared on equal footing. Inference is carried out with an efficient posterior sampling algorithm that jointly handles team abilities, tier structure, home advantage, and structural balance.
Simulation studies validate the proposed method, showing accurate recovery
of the latent competitive structure and correct performance of the
structural-balance testing procedure. Applied to the NBA and NFL, it reveals that the two leagues exhibit similar global structural balance despite very different competitive organizations: the NBA has three latent tiers, while the NFL has two broader ones, and the NFL's sparser schedule produces considerably more posterior uncertainty. Overall parity, it turns out, can mask quite different forms of competitive stratification—something standings-based measures alone cannot detect.
\end{abstract}


\section{Introduction}
\label{sec:introduction}

Competitive balance sits at the heart of professional sports leagues and has long occupied sports economists and league managers. Teams closely matched in underlying strength make for less predictable outcomes and seasons that stay meaningful into their final weeks, and that unpredictability is a large part of what keeps fans engaged — engagement that, in turn, underwrites much of a league's commercial value. Persistent gaps in team quality work the opposite way: they harden into a hierarchical structure where a handful of clubs dominate and the uncertainty of outcome that draws fans in the first place starts to erode. That's why competitive balance comes up so often in debates over league governance, talent allocation, revenue sharing, and the long-run sustainability of competition. Leagues have responded with an array of parity-promoting policies, including player drafts, salary caps, luxury taxes, revenue sharing, and schedule design, and much of the classical literature in sports economics examines these instruments in exactly this light \citep{Vrooman1995,FortQuirk1995,Szymanski2001,Szymanski2003,Zimbalist2002,FortMaxcy2003}. Measuring competitive balance well matters for a reason beyond description: it's how we tell whether these policies are actually delivering the parity they're designed to produce. The standard toolkit for measuring competitive balance, which includes the dispersion of winning percentages, concentration indices, and related measures \citep{utt2002pitfalls,owen2007measuring,Manasis2013,ManasisNtzoufras2014,Manasis2022}, summarizes realized outcomes well. But it has two real limitations. It mixes persistent differences in underlying team strength with plain game-level noise, and it says almost nothing about how teams sort into latent competitive groups or how uncertain that grouping itself is. As a result, two leagues can look equally balanced on paper while their underlying competitive organization differs substantially, and what looks like a difference in standings may just reflect how much schedule information each league happens to provide.

A related but distinct feature of league competition is the presence of \emph{competitive tiers}. Teams need not fall into clearly distinguishable positions along a complete ranking — several clubs may simply be of comparable latent strength and together form a common competitive level. That means competitive balance and competitive stratification capture different things: two leagues can show similar overall heterogeneity while organizing teams into very different numbers of tiers, and two leagues with the same number of tiers can still differ sharply in how far apart those tiers sit. We treat these as separate inferential problems.

\citet{basini2023assessing} make the value of latent grouping concrete, using a Bayesian stochastic block model to trace more than four decades of English football and showing how shifts in the number, sizes, and interaction patterns of latent groups surface features of competition that conventional balance indices miss. The natural home for this kind of question is the Bradley--Terry--Luce model \citep{bradley1952rank,Luce1959}, which represents game outcomes through differences in latent ability and underlies a large literature on sports ranking, forecasting, Bayesian inference, and scalable computation \citep{davidson1970extending,glickman1999parameter,hunter2004mm,caron2012efficient,mollica2021bayesian,tian2024spectral,pearce2025modeling}. But the standard version of that model assigns every competitor a continuous latent ability, so no two competitors ever share a strength level with positive probability — every team is distinguishable from every other whether or not that distinction means anything.

Several lines of work relax this. The ranking lasso groups Bradley--Terry effects through penalized fusion \citep{masarotto2012ranking}; a separate body of Bayesian and mixture-based work handles clustered rankings, partial orders, and related structures more generally \citep{marden1996analyzing,murphy2003mixtures,busse2007cluster,dunson2008bayesian,deng2014bayesian,li2022bayesian}. Two recent papers push this further by inferring latent rank groups directly: \citet{pearce2025bayesian} propose a rank-clustering Bradley--Terry--Luce model that jointly estimates the number and memberships of latent rank clusters, and \citet{santi2026bradley} combine paired-comparison ranking with latent block structure to recover tied competitive levels.

What none of this tells you is how far apart the tiers actually are. A league could have three tightly bunched tiers and a league could have two tiers separated by a canyon, and a method built to answer "who belongs together" has no way to distinguish those two situations — yet they represent very different competitive environments. That's the gap we're after: we infer the tiers, but we also put a number on how much separation exists between them, on a scale that lets you compare that separation across leagues regardless of how many tiers or teams each one has.

We develop a Bayesian extension of the Bradley--Terry model that studies competitive tier structure and global structural balance jointly. A mixture of finite mixtures prior \citep{miller2018mixture} is placed on team abilities, so the number of occupied tiers and their memberships are both learned from the observed games. Our work builds on recent approaches that characterize competitive structure through latent groups: Bayesian stochastic block models describe competition through the number, composition, and interaction patterns of latent blocks \citep{basini2023assessing}, spike and slab fusion priors identify competitors with tied latent ranks \citep{pearce2025bayesian}, and Bradley--Terry stochastic block models jointly infer tied rank groups, their strengths, and their memberships \citep{santi2026bradley}. These methods describe how competitors are grouped, but the grouping alone does not say how strongly the resulting competitive levels are separated. A league with several closely spaced tiers can be more balanced than one with only two tiers sitting far apart. We therefore introduce a global structural balance parameter that measures the dispersion of tier-specific abilities separately from the latent partition. The partition captures competitive stratification, while the balance parameter quantifies how large the heterogeneity among tiers actually is, which gives a common scale for comparing leagues with very different competitive structures.
Our theoretical development makes this distinction precise. We characterize the balance parameter through the exact distribution of average between-tier separation, and we show that team-level heterogeneity decomposes into the magnitude of between-tier differences and the prevalence of cross-tier pairs. Greater structural balance, we further show, corresponds to fewer practically distinguishable competitive levels at a fixed resolution, which clarifies why tier count by itself is not an adequate measure of balance. The common scale also lets us move beyond descriptive comparison. We develop Bayesian procedures for comparing structural balance between systems directly, including directional posterior probabilities and practical-equivalence assessments, even when the systems differ in their numbers of teams and comparison designs. Since leagues also differ markedly in their schedules, we study identification under incomplete comparison graphs and separate the underlying structural quantity from the amount of information available to estimate it. Posterior computation combines the MFM representation with P\'olya--Gamma augmentation \citep{polson2013bayesian} and jointly propagates uncertainty in team abilities, tier memberships, home advantage, structural balance, and cross-league comparisons.
Our application to the NBA and NFL shows why both the measure and the comparison procedure matter. Despite very different observed standings and schedule densities, the posterior analysis offers little directional evidence that either league is globally more balanced, which points to similar levels of structural balance. The latent competitive organizations are nonetheless quite different: the NBA has three competitive tiers, while the NFL has two broader ones. The NFL's much sparser schedule also leads to greater posterior uncertainty, both in league-specific inference and in the cross-league comparison. Similar overall parity, it turns out, can coexist with different forms of competitive stratification, and formal uncertainty quantification can separate genuine differences in structural balance from differences that merely reflect schedule information.

The remainder of the paper is organized as follows. Section \ref{sec:method}
introduces the mixture of finite mixtures model for team abilities, defines
the balance parameter $B$, and describes the MCMC algorithm used for
posterior sampling and post MCMC inference procedure. Section \ref{sec:theory} develops the theoretical
results behind $B$, and Section \ref{sec:testing} turns that interpretation
into a practical procedure for comparing balance across two systems.
Section \ref{sec:simulation} reports extensive simulations that check
recovery and the comparison procedure under schedules of varying density,
and Section \ref{sec:application-nba-nfl} applies the framework to the NBA
and NFL. We conclude in Section \ref{sec:conclusion} with a summary of the
findings and directions for future work.


\section{Methodology}
\label{sec:method}

\subsection{Paired comparisons and latent tier structure}
\label{sec:latent-tier-model}

Let $i=1,\ldots,n$ index observational units and
$e=1,\ldots,M$ index paired comparisons. For comparison $e$, let
$a_e$ and $b_e$ denote the two participating units, and let $Y_e=1$ if
unit $a_e$ is preferred to unit $b_e$ and $Y_e=0$ otherwise. We assume
the Bradley--Terry model \citep{bradley1952rank}
\begin{equation}
Y_e\mid\boldsymbol\theta,h
\sim
\operatorname{Bernoulli}(p_e),
\qquad
\operatorname{logit}(p_e)
=
\theta_{a_e}-\theta_{b_e}+h,
\label{eq:bt-model}
\end{equation}
where $\theta_i$ is the latent ability of unit $i$. The parameter $h$
represents a systematic first-position effect, with
$h\sim N(0,\sigma_h^2)$ when estimated and $h=0$ under a symmetric
comparison design.

The comparison structure can be represented by an oriented incidence
matrix. Let $\boldsymbol q_e\in\mathbb R^n$ contain a $1$ in position
$a_e$, a $-1$ in position $b_e$, and zeros elsewhere, and define
\[
Q_M
=
(\boldsymbol q_1,\ldots,\boldsymbol q_M)^\top.
\]
Then the linear predictor is
$\boldsymbol q_e^\top\boldsymbol\theta+h$. This representation connects
identifiability of the latent effects to the comparison graph and its
incidence matrix \citep{dong2026statistical}.

We assume that units with comparable latent abilities may share a common
tier. Let $K$ denote the nominal number of mixture components and assign
the mixture-of-finite-mixtures prior \citep{miller2018mixture}
\begin{equation}
\begin{aligned}
K
&\sim
1+\operatorname{Poisson}(\lambda_K),\\
\boldsymbol\pi\mid K
&\sim
\operatorname{Dirichlet}(\gamma,\ldots,\gamma),\\
z_i\mid\boldsymbol\pi,K
&\sim
\operatorname{Categorical}(\boldsymbol\pi),
\qquad
\theta_i=\mu_{z_i}.
\end{aligned}
\label{eq:mfm-prior}
\end{equation}

Conditional on a global precision parameter $\phi$,
\begin{equation}
\mu_k\mid\phi,K
\overset{\mathrm{iid}}{\sim}
N\left(
0,\frac{\tau_0^2}{\phi}
\right),
\qquad
k=1,\ldots,K,
\label{eq:tier-effect-prior}
\end{equation}
where $\tau_0>0$ is a fixed reference scale.

Let
\[
\mathcal P
=
\{C_1,\ldots,C_T\}
\]
denote the occupied partition after empty mixture components are removed,
with $T=|\mathcal P|$. Thus $T$ is the number of occupied tiers and
$\mathcal P$ determines their memberships. Mixture labels have no
substantive meaning and no ordering constraint is imposed during posterior
sampling.

For a fixed partition, let $Z_{\mathcal P}$ be the $n\times T$
membership matrix and define
\[
D_{\mathcal P}
=
Q_MZ_{\mathcal P}.
\]
Then
\[
\boldsymbol\theta
=
Z_{\mathcal P}\boldsymbol\mu,
\qquad
\boldsymbol\psi
=
D_{\mathcal P}\boldsymbol\mu
+
h\boldsymbol 1_M.
\]
For a connected comparison graph,
$\operatorname{rank}(D_{\mathcal P})=T-1$, so the occupied tier effects
are identifiable up to a common location. A precise rank condition is
given in Section~\ref{sec:theory-identification}.

\subsection{Structural balance}
\label{sec:balance-parameter}

We assign
\[
\phi
\sim
\operatorname{Gamma}(a_\phi,b_\phi),
\]
using the shape-rate parameterization, and define
\begin{equation}
B
=
\frac{\phi}{1+\phi},
\qquad
\phi
=
\frac{B}{1-B},
\qquad
0<B<1.
\label{eq:balance-parameter}
\end{equation}
The parameter $B$ measures the global magnitude of structural balance.
Larger values of $B$ imply greater precision and hence smaller separation
among the latent tier effects.

For two distinct tiers $c$ and $d$,
\begin{equation*}
\mu_c-\mu_d\mid B
\sim
N\left(
0,
2\tau_0^2\frac{1-B}{B}
\right),
\label{eq:tier-contrast}
\end{equation*}
and therefore
\[
E\left\{
(\mu_c-\mu_d)^2
\mid B
\right\}
=
2\tau_0^2\frac{1-B}{B}.
\]

For $T\geq2$, define the average squared separation
\begin{equation}
D_T
=
\frac{2}{T(T-1)}
\sum_{c<d}
(\mu_c-\mu_d)^2.
\label{eq:tier-dispersion}
\end{equation}
Then
\begin{equation}
E(D_T\mid B,T)
=
2\tau_0^2\frac{1-B}{B},
\label{eq:expected-tier-dispersion}
\end{equation}
which does not depend on $T$. Thus the latent partition
$(T,\mathcal P)$ describes the tier structure, whereas $B$ measures the
magnitude of between-tier heterogeneity.

The interpretation also carries to the paired-comparison scale. In the
absence of a position effect,
\[
p_{cd}
=
\operatorname{expit}(\mu_c-\mu_d).
\]
As $B\rightarrow1$, the tier contrasts shrink toward zero and
$p_{cd}\rightarrow1/2$. Hence larger values of $B$ correspond to greater
structural balance.


\subsection{Posterior computation}
\label{sec:computation}

We develop a semi-collapsed Gibbs sampler \citep{neal2000markov} that combines the MFM partition
representation with P\'olya--Gamma augmentation
\citep{polson2013bayesian}. The mixture weights and nominal component count
are integrated out when updating the occupied partition, while the effect of
a proposed new tier is integrated analytically. The sampler therefore avoids
reversible-jump moves and does not require the number of occupied tiers to be
fixed in advance. Full derivations of the sampling updates are provided in
the Supporting Materials.

For the current partition, define
\[
\psi_e
=
\boldsymbol d_e^\top\boldsymbol\mu+h,
\qquad
\kappa_e
=
Y_e-\frac12,
\]
where $\boldsymbol d_e^\top$ is row $e$ of $D_{\mathcal P}$. Introduce
the P\'olya--Gamma latent variables
\begin{equation}
\omega_e\mid\psi_e
\sim
\operatorname{PG}(1,\psi_e),
\qquad
e=1,\ldots,M.
\label{eq:pg-augmentation}
\end{equation}
Let
\[
\Omega
=
\operatorname{diag}(\omega_1,\ldots,\omega_M),
\qquad
\boldsymbol\kappa
=
(\kappa_1,\ldots,\kappa_M)^\top.
\]

Conditional on $\boldsymbol\omega$,
\begin{equation}
\begin{pmatrix}
\boldsymbol\mu\\
h
\end{pmatrix}
\Bigm|\mathrm{rest}
\sim
N\left(
\mathcal Q^{-1}\boldsymbol b,
\mathcal Q^{-1}
\right),
\label{eq:gaussian-full-conditional}
\end{equation}
where
\[
\mathcal Q
=
\begin{pmatrix}
D_{\mathcal P}^{\top}\Omega D_{\mathcal P}
+\phi\tau_0^{-2}I_T
&
D_{\mathcal P}^{\top}\Omega\boldsymbol 1_M
\\
\boldsymbol 1_M^{\top}\Omega D_{\mathcal P}
&
\boldsymbol 1_M^{\top}\Omega\boldsymbol 1_M+\sigma_h^{-2}
\end{pmatrix},
\]
and
\[
\boldsymbol b
=
\begin{pmatrix}
D_{\mathcal P}^{\top}\boldsymbol\kappa\\
\boldsymbol 1_M^{\top}\boldsymbol\kappa
\end{pmatrix}.
\]
When $h=0$, the last row and column are omitted.

Integrating out $(K,\boldsymbol\pi)$ gives
\begin{equation}
p(\mathcal P)
=
V_n(T)
\prod_{c=1}^{T}
\gamma^{(n_c)},
\qquad
V_n(T)
=
\sum_{k=T}^{\infty}
\frac{k_{[T]}}
{(\gamma k)^{(n)}}
p_K(k),
\label{eq:mfm-partition}
\end{equation}
where $n_c=|C_c|$, $x^{(r)}$ denotes a rising factorial, and $x_{[r]}$
denotes a falling factorial.

To update $z_i$, remove unit $i$ from its current tier and let $t$ denote
the number of remaining occupied tiers. For an existing tier $c$,
\begin{equation}
W_{ic}
=
(n_{-i,c}+\gamma)
\exp\{\ell_i(c)\},
\label{eq:existing-tier-weight}
\end{equation}
where $\ell_i(c)$ is the P\'olya--Gamma augmented log-likelihood
contribution from comparisons involving unit $i$.

For a proposed new tier, define
\[
A_i
=
\frac{\phi}{\tau_0^2}
+
\sum_{e\in\mathcal E_i}\omega_e,
\qquad
C_i
=
\sum_{e\in\mathcal E_i}
s_{ie}
(\kappa_e-\omega_e r_{ie}),
\]
and
\[
H_i
=
\sum_{e\in\mathcal E_i}
\left(
\kappa_e r_{ie}
-
\frac{\omega_e}{2}r_{ie}^2
\right).
\]
The integrated contribution of the proposed tier is
\[
m_i^{\mathrm{new}}
=
\exp(H_i)
\left(
\frac{\phi/\tau_0^2}{A_i}
\right)^{1/2}
\exp\left(
\frac{C_i^2}{2A_i}
\right),
\]
giving
\begin{equation}
W_i^{\mathrm{new}}
=
\gamma
\frac{V_n(t+1)}{V_n(t)}
m_i^{\mathrm{new}}.
\label{eq:new-tier-weight}
\end{equation}
If a new tier is selected,
\begin{equation}
\mu_{\mathrm{new}}
\mid\mathrm{rest}
\sim
N\left(
\frac{C_i}{A_i},
\frac{1}{A_i}
\right).
\label{eq:new-tier-effect}
\end{equation}

Conditional on the occupied tier effects,
\begin{equation}
\phi\mid\boldsymbol\mu,T
\sim
\operatorname{Gamma}
\left(
a_\phi+\frac{T}{2},
\;
b_\phi+
\frac{1}{2\tau_0^2}
\sum_{c=1}^{T}\mu_c^2
\right),
\label{eq:phi-full-conditional}
\end{equation}
after which $B=\phi/(1+\phi)$.

Algorithm~\ref{alg:gibbs} summarizes the sampler.

\begin{algorithm}[t]
\caption{Semi-collapsed Gibbs sampler}
\label{alg:gibbs}
\begin{algorithmic}[1]

\State Initialize
$\mathcal P^{(0)}$,
$\boldsymbol\mu^{(0)}$,
$h^{(0)}$, and $\phi^{(0)}$.

\For{$s=1,\ldots,S$}

    \State Compute
    $\psi_e=\boldsymbol d_e^\top\boldsymbol\mu+h$
    and sample $\omega_e$ according to
    \eqref{eq:pg-augmentation}.

    \For{$i=1,\ldots,n$}

        \State Remove unit $i$ from its current tier and delete the tier if
        it becomes empty.

        \State Compute the existing-tier weights in
        \eqref{eq:existing-tier-weight} and the new-tier weight in
        \eqref{eq:new-tier-weight}.

        \State Sample $z_i$ from the normalized allocation weights.

        \If{a new tier is selected}
            \State Sample $\mu_{\mathrm{new}}$ from
            \eqref{eq:new-tier-effect}.
        \EndIf

    \EndFor

    \State Construct the updated tier design matrix $D_{\mathcal P}$.

    \State Sample $(\boldsymbol\mu,h)$ from
    \eqref{eq:gaussian-full-conditional}.

    \State Sample $\phi$ from \eqref{eq:phi-full-conditional} and set
    $B=\phi/(1+\phi)$.

    \State Store
    $\mathcal P$, $T$, $\boldsymbol\mu$, $h$, $\phi$, and $B$.

\EndFor

\end{algorithmic}
\end{algorithm}

\subsection{Posterior summaries}
\label{sec:posterior-summary}

The MCMC output supports inference on everything the model estimates: the
number of occupied tiers, their memberships, the tier effects, and the
structural balance parameter. The posterior distribution of the occupied
tier count is estimated by
\[
\widehat{\Pr}(T=t\mid\boldsymbol Y)
=
\frac{1}{S}
\sum_{s=1}^{S}
\mathbf 1\{T^{(s)}=t\}.
\]

Because any single draw of the partition can be misleading, we summarize
partition uncertainty through the posterior similarity matrix
\[
\widehat P_{ij}
=
\frac{1}{S}
\sum_{s=1}^{S}
\mathbf 1
\left\{
z_i^{(s)}=z_j^{(s)}
\right\},
\]
so that $\widehat P_{ij}$ is the posterior probability that units $i$ and
$j$ belong to the same tier.

When a single representative partition is needed, we use the least-squares
criterion of \citet{dahl2006model}. Let
\[
A_{ij}^{(s)}
=
\mathbf 1
\left\{
z_i^{(s)}=z_j^{(s)}
\right\}
\]
be the co-clustering matrix from posterior draw $s$. The Dahl partition is
\begin{equation}
\widehat{\mathcal P}_{\mathrm{Dahl}}
=
\mathcal P^{(s^\ast)},
\qquad
s^\ast
=
\arg\min_{1\leq s\leq S}
\sum_{i<j}
\left(
A_{ij}^{(s)}-\widehat P_{ij}
\right)^2.
\label{eq:dahl-partition}
\end{equation}
Since the criterion involves only pairwise co-clustering indicators, it is
invariant to mixture labels and no post-processing for label switching is
required.

For interpretation, the tiers in
$\widehat{\mathcal P}_{\mathrm{Dahl}}$ are ordered by their posterior tier
effects. Structural balance is summarized by the posterior distribution of
$B$, with its posterior median and credible interval as the primary
summaries. When several paired-comparison populations are analyzed together,
the posterior draws of $B$ feed directly into the Bayesian comparisons of
Section~\ref{sec:testing}.


\section{Theoretical Properties}
\label{sec:theory}

This section studies two theoretical properties of the proposed framework.
We first consider identifiability under incomplete comparison designs and
then provide a structural interpretation of the balance parameter. These
results clarify when the model can learn team differences from an incomplete
schedule and how the balance parameter can be interpreted in practical
comparisons of competitive structure. Proofs of all results are provided in
the Supporting Materials.

\subsection{Identification under incomplete comparison designs}
\label{sec:theory-identification}

The Bradley--Terry likelihood involves the latent abilities only through
pairwise differences. Therefore, shifting all tier effects by a common
constant leaves the likelihood unchanged. The common location is not
identified, but under a connected comparison design all contrasts among
occupied tiers are identifiable.

\begin{proposition}[Graph identification of tier contrasts]
\label{prop:graph-identification}

Suppose the observed comparison graph is connected and
$\mathcal P=\{C_1,\ldots,C_T\}$ contains $T$ nonempty blocks. Then
\[
\operatorname{rank}(D_{\mathcal P})
=
T-1,
\]
or equivalently,
\[
\operatorname{null}(D_{\mathcal P})
=
\operatorname{span}\{\boldsymbol 1_T\}.
\]
Thus all contrasts among the occupied tier effects are identifiable.

If the first-position effect $h$ is included, let $H_{\mathcal P}$ be a
$T\times(T-1)$ contrast matrix satisfying
\[
H_{\mathcal P}^{\top}\boldsymbol 1_T=0,
\qquad
H_{\mathcal P}^{\top}H_{\mathcal P}=I_{T-1}.
\]
Then $(\boldsymbol\beta_{\mathcal P},h)$ is identifiable whenever
\[
\operatorname{rank}
\left[
D_{\mathcal P}H_{\mathcal P},
\boldsymbol 1_M
\right]
=
T.
\]
\end{proposition}

Proposition~\ref{prop:graph-identification} distinguishes identifiability
from estimation precision. Connectivity ensures that occupied tier contrasts
are identifiable, whereas the number and distribution of observed comparisons
determine how precisely these contrasts can be estimated.

To characterize estimation under increasing comparison information, suppose
the true occupied partition is $\mathcal P^\ast$ and consider a fixed-$n$
regime with $M\rightarrow\infty$. Assume that the limiting comparison graph
is connected, the true comparison probabilities are bounded away from zero
and one, and
\begin{equation}
\frac{1}{M}
X_{\mathcal P^\ast,M}^{\top}
W_M^\ast
X_{\mathcal P^\ast,M}
\longrightarrow
\mathcal I^\ast,
\label{eq:limiting-information}
\end{equation}
where $\mathcal I^\ast$ is positive definite.

\begin{proposition}[Conditional posterior concentration]
\label{prop:conditional-contraction}

Suppose the prior density for the identifiable tier contrasts and $h$ is
positive and continuous in a neighborhood of the true parameter. Conditional
on $\mathcal P=\mathcal P^\ast$, for any sequence
$r_M\rightarrow\infty$,
\begin{equation*}
\Pi
\left[
\left\|
\boldsymbol\vartheta-\boldsymbol\vartheta^\ast
\right\|
>
\frac{r_M}{\sqrt M}
\;\middle|\;
Y^{(M)},\mathcal P^\ast
\right]
\longrightarrow0
\label{eq:conditional-contraction}
\end{equation*}
in $P^\ast$-probability.
\end{proposition}

Proposition~\ref{prop:conditional-contraction} shows that, conditional on the
occupied partition, increasing the number of informative comparisons improves
estimation of the tier contrasts at the usual parametric rate. Thus the amount
of schedule information affects estimation precision but does not change the
definition of structural balance.

\subsection{Structural interpretation of competitive balance}
\label{sec:B-theory}

From \eqref{eq:balance-parameter},
\[
\phi
=
\frac{B}{1-B},
\qquad
0<B<1,
\]
and the occupied tier effects satisfy
\[
\mu_c\mid B
\overset{\mathrm{iid}}{\sim}
N\left(
0,
\tau_0^2\frac{1-B}{B}
\right).
\]
Therefore, for two distinct tiers,
\[
\mu_c-\mu_d\mid B
\sim
N\left(
0,
2\tau_0^2\frac{1-B}{B}
\right).
\]

Recall the average squared separation $D_T$ defined in
\eqref{eq:tier-dispersion}.

\begin{theorem}[Exact distribution of tier dispersion]
\label{thm:B-dispersion}

Conditional on $B$ and $T\geq2$,
\begin{equation}
\frac{(T-1)B}
{2\tau_0^2(1-B)}
D_T
\sim
\chi^2_{T-1}.
\label{eq:exact-tier-dispersion}
\end{equation}
Consequently,
\[
E(D_T\mid B,T)
=
2\tau_0^2
\frac{1-B}{B},
\]
and
\[
\operatorname{Var}(D_T\mid B,T)
=
\frac{8\tau_0^4}{T-1}
\left(
\frac{1-B}{B}
\right)^2.
\]

Moreover, if $0<B_2<B_1<1$, then
\[
|\mu_c-\mu_d|
\mid B_1
\leq_{\mathrm{st}}
|\mu_c-\mu_d|
\mid B_2.
\]
\end{theorem}

Theorem~\ref{thm:B-dispersion} provides a direct interpretation of $B$.
The expected average between-tier separation does not depend on the number of
occupied tiers, the number of competitors, or the number of observed
comparisons. Therefore, $B$ provides a common measure of between-tier
heterogeneity across different competitive systems.

To connect this quantity with team-level heterogeneity, define
\[
D_n^{\mathrm{unit}}
=
\frac{2}{n(n-1)}
\sum_{1\leq i<j\leq n}
(\theta_i-\theta_j)^2
\]
and
\[
q(\mathcal P)
=
1-
\frac{\sum_{c=1}^{T}n_c(n_c-1)}
{n(n-1)}.
\]

\begin{corollary}[Population-level dispersion decomposition]
\label{cor:unit-dispersion}

Conditional on $B$ and $\mathcal P$,
\begin{equation*}
E
\left(
D_n^{\mathrm{unit}}
\mid
B,\mathcal P
\right)
=
q(\mathcal P)
\,2\tau_0^2
\frac{1-B}{B}.
\label{eq:unit-dispersion-decomposition}
\end{equation*}
\end{corollary}

Corollary~\ref{cor:unit-dispersion} shows that overall team-level
heterogeneity depends on two features: the magnitude of separation between
occupied tiers, measured by $2\tau_0^2(1-B)/B$, and the proportion of team
pairs belonging to different tiers, measured by $q(\mathcal P)$. Therefore,
two systems may have similar structural balance while exhibiting different
levels of overall heterogeneity because their tier compositions differ.

\subsection{Structural balance and effective tier complexity}
\label{sec:effective-tier-complexity}

The occupied tier count $T$ is learned from the posterior partition and is not
determined by $B$. However, $B$ is related to the number of competitive levels
that remain distinguishable at a given practical resolution.

Fix $\delta>0$ and define a threshold graph on the occupied tiers by connecting
$c$ and $d$ whenever
\[
|\mu_c-\mu_d|
\leq
\delta.
\]
Let $T_\delta(B)$ denote the number of connected components of this graph.

\begin{proposition}[Balance and effective tier complexity]
\label{prop:effective-tier-count}

Fix $T\geq2$ and $\delta>0$, and write
\[
\mu_c(B)
=
\tau_0
\left(
\frac{1-B}{B}
\right)^{1/2}
Z_c,
\qquad
Z_c
\overset{\mathrm{iid}}{\sim}
N(0,1).
\]
Under this common-$Z$ coupling, if $B_1>B_2$, then
\begin{equation*}
T_\delta(B_1)
\leq
T_\delta(B_2)
\qquad
\text{almost surely},
\label{eq:effective-tier-monotonicity}
\end{equation*}
and consequently
\[
E\{T_\delta(B_1)\mid T\}
\leq
E\{T_\delta(B_2)\mid T\}.
\]

For any two distinct occupied tiers,
\[
\Pr
\left(
|\mu_c-\mu_d|>\delta
\mid B
\right)
=
2\Phi
\left[
-\frac{\delta}{\sqrt{2}\tau_0}
\left(
\frac{B}{1-B}
\right)^{1/2}
\right],
\]
which is strictly decreasing in $B$.
\end{proposition}

Proposition~\ref{prop:effective-tier-count} shows that greater structural
balance leads to smaller differences among occupied tiers and therefore fewer
practically distinguishable competitive levels at a fixed resolution
$\delta$. This result concerns effective tier complexity and does not imply
that the raw occupied MFM tier count $T$ must decrease monotonically with $B$
in every finite data set.

Finally, for a neutral comparison between tiers $c$ and $d$, let
\[
p_{cd}
=
\operatorname{expit}(\mu_c-\mu_d).
\]
Then
\[
E
\left[
\left\{
\operatorname{logit}(p_{cd})
\right\}^2
\mid B
\right]
=
2\tau_0^2
\frac{1-B}{B},
\]
so larger values of $B$ correspond to smaller variation in pairwise
log-odds contrasts.


\section{Bayesian Comparison of Structural Balance Across Systems}
\label{sec:testing}

The results in Section~\ref{sec:B-theory} provide a common scale for comparing
competitive balance across paired-comparison systems. Consider two systems
indexed by $\ell=1,2$, with balance parameters $B_1$ and $B_2$. The systems
may differ in the number of competitors, occupied tiers, and observed
comparisons, but are analyzed using the same reference scale $\tau_0$.

By Theorem~\ref{thm:B-dispersion}, define
\begin{equation*}
d(B)
=
E(D_T\mid B,T)
=
2\tau_0^2\frac{1-B}{B}.
\label{eq:structural-dispersion}
\end{equation*}
Because $d(B)$ is strictly decreasing on $(0,1)$, larger values of $B$
correspond to smaller between-tier separation and therefore greater structural
balance. Importantly, $d(B)$ does not depend on the number of competitors,
the number of occupied tiers, or the number of observed comparisons. These
features affect posterior uncertainty and the realized competitive structure,
but not the interpretation of structural balance. Therefore, the same scale
can be used to compare systems with different competitive structures and
schedules.

\subsection{Directional comparison of structural balance}
\label{sec:balance-comparison}

Define
\begin{equation*}
\Delta_B
=
B_1-B_2.
\label{eq:delta-B}
\end{equation*}
Because $d(B)$ is strictly decreasing,
\[
\Delta_B>0
\quad\Longleftrightarrow\quad
d(B_1)<d(B_2).
\]
Thus $\Delta_B>0$ indicates greater structural balance in system 1, whereas
$\Delta_B<0$ indicates greater structural balance in system 2.

Given posterior draws,
\[
\Delta_B^{(s)}
=
B_1^{(s)}-B_2^{(s)}.
\]
We summarize the posterior distribution of $\Delta_B$ using its median,
credible interval, and the directional probability
\begin{equation*}
P_{12}
=
\Pr(B_1>B_2\mid Y_1,Y_2).
\label{eq:directional-probability}
\end{equation*}
Values near one provide evidence that system 1 is more balanced, values near
zero favor system 2, and values near one half indicate little directional
evidence.

\subsection{Practical differences in structural balance}
\label{sec:bayesian-equivalence}

A directional comparison does not indicate whether the difference in
structural balance is practically meaningful. We therefore define
\begin{equation*}
\Delta_d
=
d(B_2)-d(B_1),
\label{eq:delta-d}
\end{equation*}
so that
\[
\Delta_d>0
\quad\Longleftrightarrow\quad
B_1>B_2.
\]

For a prespecified practical margin $\epsilon_d>0$, consider
\begin{equation*}
H_-:\Delta_d<-\epsilon_d,
\qquad
H_0:|\Delta_d|\leq\epsilon_d,
\qquad
H_+:\Delta_d>\epsilon_d.
\label{eq:testing-regions}
\end{equation*}
Here, $H_-$ represents meaningfully greater balance in system 2,
$H_0$ represents practical similarity, and $H_+$ represents meaningfully
greater balance in system 1.

The corresponding posterior probabilities are
\begin{equation*}
\begin{aligned}
\pi_-
&=
\Pr(\Delta_d<-\epsilon_d\mid Y_1,Y_2),\\
\pi_0
&=
\Pr(|\Delta_d|\leq\epsilon_d\mid Y_1,Y_2),\\
\pi_+
&=
\Pr(\Delta_d>\epsilon_d\mid Y_1,Y_2),
\end{aligned}
\label{eq:testing-probabilities}
\end{equation*}
where
\[
\pi_-+\pi_0+\pi_+=1.
\]
The vector $(\pi_-,\pi_0,\pi_+)$ distinguishes evidence for a meaningful
difference from evidence for practical similarity. In particular, a
directional probability near one half does not imply practical equivalence.

For posterior draws,
\[
\Delta_d^{(s)}
=
2\tau_0^2
\left\{
\frac{1-B_2^{(s)}}{B_2^{(s)}}
-
\frac{1-B_1^{(s)}}{B_1^{(s)}}
\right\}.
\]

For calibration, the root mean squared tier separation is
\begin{equation*}
s(B)
=
\{d(B)\}^{1/2}
=
\sqrt{2}\tau_0
\left(
\frac{1-B}{B}
\right)^{1/2},
\label{eq:rms-tier-separation}
\end{equation*}
which is expressed on the latent Bradley--Terry log-odds scale. This provides
a direct way to specify $\epsilon_d$ in terms of a practically meaningful
difference in tier separation.

Evidence for practical similarity can also be summarized using the interval
Bayes factor
\begin{equation*}
BF_{01}^{(\epsilon)}
=
\frac{
\Pr(H_0\mid Y_1,Y_2)/
\Pr(H_1\mid Y_1,Y_2)
}{
\Pr(H_0)/
\Pr(H_1)
},
\qquad
H_1:|\Delta_d|>\epsilon_d.
\label{eq:interval-bayes-factor}
\end{equation*}
Values greater than one indicate increased support for practical similarity
relative to the prior, whereas values below one favor a meaningful difference.
We report this quantity as a complementary posterior summary together with
$(\pi_-,\pi_0,\pi_+)$.

\section{Simulation study}
\label{sec:simulation}

We conducted a factorial simulation study to evaluate the proposed framework
under comparison structures and competitive configurations motivated by
professional sports applications. The simulations address two related
questions. First, we examine how accurately the model recovers latent team
abilities, competitive tiers, and structural balance under schedules with
different amounts of comparison information. Second, we evaluate whether the
proposed posterior comparison of structural balance remains informative when
two systems differ substantially in their numbers of teams and observed
comparisons.

\subsection{Estimation and latent-tier recovery}
\label{sec:simulation-estimation}

The simulations use $n=30$ teams under three comparison-graph designs. The
first is the observed NBA regular-season schedule with 1,230 games. To see
what happens as comparison information is stripped away while keeping the
graph connected, we add a complete double round-robin with 870 games and a
complete single round robin with 435 games. The three designs step down in
informativeness, which lets us watch how schedule density affects recovery of
the latent competitive structure.

Each design is crossed with two structural settings,
\[
(B,T)=(0.2,4)
\qquad\text{and}\qquad
(B,T)=(0.5,3),
\]
a less balanced system with four tiers and a more balanced one with three.
The second setting is deliberately the harder one for tier recovery: since
larger $B$ means smaller between-tier separation, its tiers sit closer
together even though there are fewer of them. Within each setting we try
both balanced and unbalanced tier allocations,
\[
(8,8,7,7)
\qquad\text{and}\qquad
(3,12,12,3)
\]
for four tiers, and
\[
(10,10,10)
\qquad\text{and}\qquad
(5,20,5)
\]
for three. The unbalanced layouts test whether recovery suffers when most
teams pile into one or two levels instead of spreading across tiers.

Team labels are randomly assigned to tiers in each replicate. With
$\tau_0=1$, the tier abilities are scaled so that the implied structural
balance matches the target exactly, and games are generated from
\begin{equation*}
\Pr(Y_{ij}=1\mid\theta,h)
=
\operatorname{logit}^{-1}(\theta_i-\theta_j+h),
\qquad h=0.35.
\end{equation*}

We ran 100 independent replicates for each of the 12 simulation settings.
For each replicate, the MFM--BT model was fitted using two independently
initialized MCMC chains. Each chain consisted of 6,000 iterations, with the
first 1,500 iterations discarded as burn-in and every third subsequent
iteration retained, yielding 1,500 posterior draws per chain. We used
$\lambda_K=1$, $\tau_0=1$, and $\gamma=1$ throughout the simulation study. The number of occupied tiers is
inferred from the data rather than fixed at the truth. Clustering accuracy
is measured by the adjusted Rand index (ARI) \citep{hubert1985comparing} of the Dahl representative
partition, and tier-count recovery by
\[
100R^{-1}\sum_{r=1}^{R}
I(\widehat T_r=T).
\]
Ability estimation is judged by mean squared error (MSE), ranking by the
Spearman correlation between estimated and true ability ranks, and the
system-level parameters by $B$-MSE and $h$-MSE from their posterior means.
These criteria deliberately separate recovery of the discrete tier structure
from estimation of the continuous balance parameter.

We compare MFM--BT with three two-stage benchmark procedures: BT--GMM,
which clusters ordinary Bradley--Terry ability estimates using a Gaussian
mixture model \citep{mclachlan2000finite}; BT--DP, which applies a
DP-means clustering procedure to the Bradley--Terry estimates
\citep{kulis2012revisiting}; and BT--pairwise-fusion, which uses an
all-pairs $L_1$ fusion penalty motivated by the ranking-lasso method
\citep{masarotto2012ranking}, all fitted
to exactly the same graphs and outcomes. Each benchmark selects its own
number of clusters from its fitted model, since in practice neither the
number nor the composition of competitive tiers is known in advance.

\begin{figure}[ht]
\centering
\includegraphics[width=\textwidth]{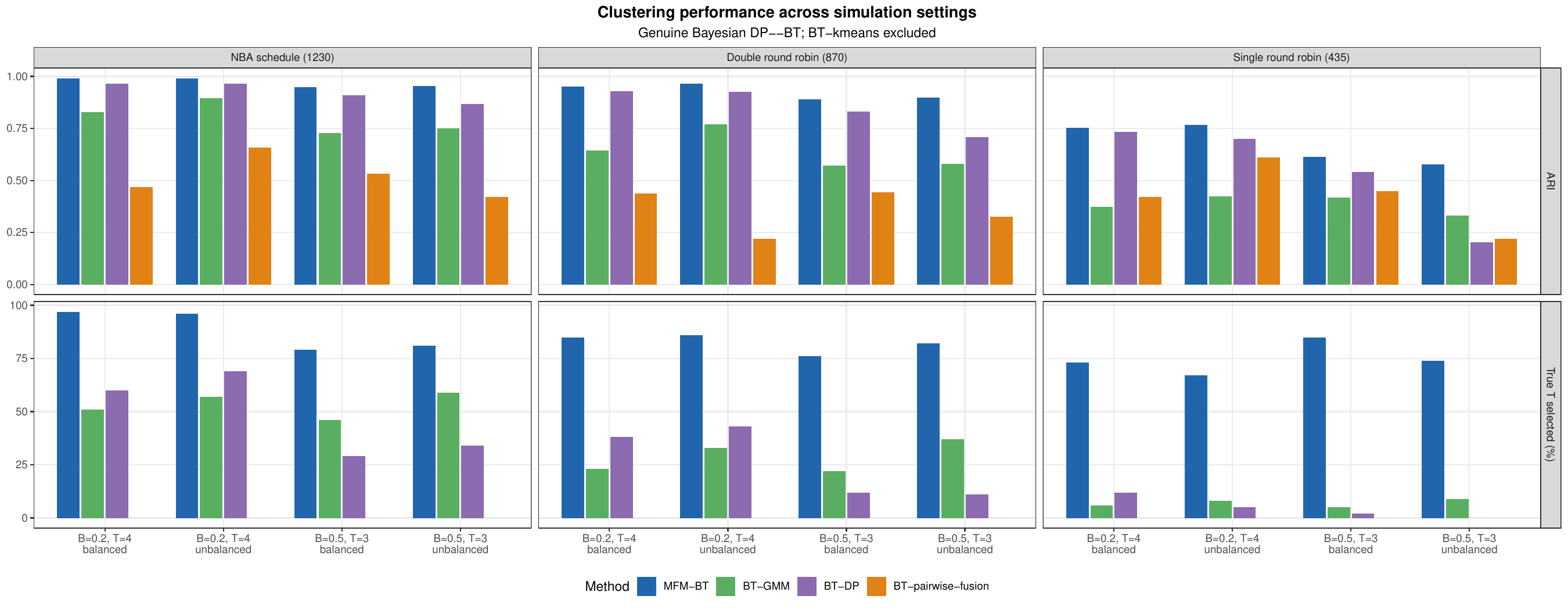}
\caption{Clustering performance across the 12 simulation designs.}
\label{fig:clustering-results}
\end{figure}

Figure~\ref{fig:clustering-results} summarizes clustering recovery across the
12 simulation settings. The upper row reports the adjusted Rand index (ARI),
and the lower row reports the percentage of replicates that recover the true
number of tiers. Within each graph design, the four bars correspond to the
two $(B,T)$ settings under balanced and unbalanced tier allocations. For
MFM--BT and BT--DP, the reported partitions are obtained from the
Dahl-selected posterior representative partition.

MFM--BT achieves the best clustering performance across nearly all settings.
Its advantage is especially clear for the sparse single-round-robin design and
the more difficult $(B,T)=(0.5,3)$ setting, where the tier differences are
smaller. BT--DP is generally the strongest benchmark in terms
of ARI, but it tends to overestimate the number of tiers. On average, it
estimates approximately 5.13 tiers when the true number is either 3 or 4,
which explains its relatively low true-tier recovery rate. 
BT--GMM performs moderately well under the denser designs but loses accuracy
as comparison information decreases. It tends to merge some of the true tiers,
leading to fewer estimated tiers than the truth. BT--pairwise-fusion has the
lowest ARI and consistently underestimates the number of tiers, recovering the
true tier count in none of the simulation replicates.

Table~\ref{tab:simulation-overall} summarizes performance across the 12
simulation settings, with 100 replicates for each setting. MFM--BT achieves
the highest average ARI, 0.858, and recovers the true number of tiers in
81.8\% of replicates. The genuine Bayesian BT--DP benchmark performs second,
with an average ARI of 0.774 and a true-tier recovery rate of 26.2\%.
BT--GMM recovers the true tier count in 29.7\% of replicates, while
BT--pairwise-fusion never recovers the true count. MFM--BT also has the
smallest ability MSE and the highest rank correlation. Thus, its advantage is
not limited to partition recovery: jointly modeling tier memberships and
abilities also improves continuous ability estimation and the induced team
ranking.

\begin{table}[ht]
\centering
\caption{Average performance over the 12 simulation settings and 100
replicates per setting. Entries are means with standard deviations in
parentheses.}
\label{tab:simulation-overall}
\begin{tabular}{lrrrr}
\hline
Method & ARI & True $T$ (\%) & Ability MSE & Rank correlation \\
\hline
MFM--BT
& 0.858 (0.182)
& 81.8 (38.6)
& 0.137 (0.162)
& 0.959 (0.067) \\

BT--GMM
& 0.610 (0.290)
& 29.7 (45.7)
& 0.378 (1.811)
& 0.915 (0.054) \\

BT--DP
& 0.774 (0.250)
& 26.2 (44.0)
& 0.144 (0.164)
& 0.952 (0.069) \\

BT--pairwise-fusion
& 0.433 (0.168)
& 0.0 (0.0)
& 1.597 (1.136)
& 0.913 (0.059) \\
\hline
\end{tabular}
\end{table}
This gap between methods matters for the application we have in mind. A
competitive-balance study needs more than a ranking of teams; it needs a
stable account of which teams are practically indistinguishable in latent
strength, and the much higher ARI of MFM-BT says its posterior partition
captures that organization far more faithfully. At the same time, the strong
rank correlations across several methods remind us that producing a ranking
is much easier than recovering the tier structure beneath it, so high
ranking accuracy on its own says little about stratification.

Table~\ref{tab:simulation-mfm} shows the MFM-BT estimation results by design.
Under the NBA schedule recovery is excellent, with ARI between 0.948 and
0.991 and true tier counts recovered in 79\% to 97\% of replicates.
Performance slips moderately under the double round robin and more sharply
under the single round robin, worst of all at $(B,T)=(0.5,3)$, where ARI
falls to 0.577--0.614. The pattern is exactly what the structure predicts:
$B=0.5$ puts the tiers closer together, and with fewer games it becomes
harder to say which side of a boundary a team belongs on.

\begin{table}[ht]
\centering
\caption{MFM-BT performance by graph design, structural setting, and tier allocation.}
\label{tab:simulation-mfm}
\resizebox{\textwidth}{!}{%
\begin{tabular}{lllrrrrr}
\hline
Graph design & $(B,T)$ & Allocation & ARI & True $T$ (\%) &
Ability MSE & $B$-MSE & $h$-MSE \\
\hline
NBA & (0.2,4) & Balanced & 0.990 & 97 & 0.058 & 0.002 & 0.007 \\
NBA & (0.2,4) & Unbalanced & 0.991 & 96 & 0.036 & 0.003 & 0.004 \\
NBA & (0.5,3) & Balanced & 0.948 & 79 & 0.032 & 0.005 & 0.004 \\
NBA & (0.5,3) & Unbalanced & 0.955 & 81 & 0.026 & 0.004 & 0.004 \\
Double RR & (0.2,4) & Balanced & 0.951 & 85 & 0.104 & 0.003 & 0.012 \\
Double RR & (0.2,4) & Unbalanced & 0.964 & 86 & 0.082 & 0.004 & 0.009 \\
Double RR & (0.5,3) & Balanced & 0.891 & 76 & 0.064 & 0.005 & 0.007 \\
Double RR & (0.5,3) & Unbalanced & 0.898 & 82 & 0.050 & 0.004 & 0.006 \\
Single RR & (0.2,4) & Balanced & 0.752 & 73 & 0.425 & 0.003 & 0.025 \\
Single RR & (0.2,4) & Unbalanced & 0.768 & 67 & 0.329 & 0.010 & 0.015 \\
Single RR & (0.5,3) & Balanced & 0.614 & 85 & 0.229 & 0.006 & 0.014 \\
Single RR & (0.5,3) & Unbalanced & 0.577 & 74 & 0.209 & 0.003 & 0.010 \\
\hline
\end{tabular}}
\end{table}

The balance parameter is far more stable than individual tier memberships.
$B$-MSE stays small in nearly every design even where ARI drops
substantially: under the single-round-robin settings with $(B,T)=(0.5,3)$,
ARI sits below 0.62 while $B$-MSE remains between 0.003 and 0.006. $B$
summarizes the overall scale of heterogeneity rather than the fate of any
one team, so it doesn't need every team to land in its exact true tier to
stay well estimated.

Losing games mainly hurts local structure. Ability MSE rises noticeably as
schedules thin out, and tier assignments near close levels become
ambiguous, while the global balance signal survives because it pools
information across the whole league -- exactly the separation between
partition uncertainty and $B$-uncertainty the framework is built around.

Unbalanced tier sizes cause no systematic damage, either. In several
settings the unbalanced allocation does as well as, or slightly better
than, the balanced one, so the method does not rely on roughly equal
groups -- a good sign for sports data, where tiers routinely contain very
different numbers of teams.

Taken together, the simulations bear out the separate roles of the
partition and the balance parameter. With an informative schedule, MFM-BT
recovers abilities, rankings, and tiers accurately; as comparisons grow
scarce, uncertainty about individual abilities and memberships rises,
especially when tiers sit close together, while $B$ remains comparatively
steady. $B$ tracks the magnitude of structural heterogeneity; the
comparison graph only governs how precisely that structure can be learned
from the games at hand -- which is exactly the division of labor the theory
predicts.
\subsection{Validation of structural-balance comparisons}
\label{sec:simulation-testing}

We separately evaluated the proposed posterior procedure for comparing
structural balance across two systems. The purpose of this experiment is to
determine whether comparisons on the common $B$-scale remain meaningful when
the systems being compared differ substantially in both population size and
schedule density.

The first design was a matched comparison in which both systems contained 30
teams and used the same NBA schedule with 1,230 comparisons. The second was a
heterogeneous comparison in which system 1 used the NBA design
$(n_1=30,M_1=1230)$, while system 2 used the NFL design
$(n_2=32,M_2=272)$. The heterogeneous setting is particularly relevant to
the empirical analysis because it separates differences in the underlying
balance parameter from differences in the amount of information available to
estimate it.

For every replicate, system 1 was generated with
$(B_1,T_1)=(0.5,3)$, while system 2 was generated with
$(B_2,T_2)=(0.2,4)$. Because $B_1>B_2$, system 1 is more structurally
balanced under the data-generating model. We generated 100 replicates for each
design and fitted the same MFM-BT model independently to the two systems.
Posterior draws were used to calculate
\[
\Delta_B=B_1-B_2,
\qquad
\Delta_d=d(B_2)-d(B_1),
\qquad
d(B)=2\tau_0^2\frac{1-B}{B}.
\]
Positive values of either contrast favor system 1. We summarized the
directional probability $\Pr(B_1>B_2\mid Y)$, the corresponding probability
on the dispersion scale, and the three practical-testing probabilities
$(\pi_-,\pi_0,\pi_+)$. The practical-equivalence margin was set to
$\epsilon_d=0.4$ on the dispersion scale.

The results are summarized in Table~\ref{tab:testing-simulation}. For the
matched design, the average posterior probability that system 1 was more
balanced was 0.864, compared with 0.829 for the heterogeneous NBA--NFL
comparison. The corresponding probabilities that system 1 was more balanced
by more than the practical-equivalence margin were 0.836 and 0.795. The
posterior median dispersion contrasts were 3.46 and 3.15, respectively.
Hence, the procedure identifies the correct direction of the balance
difference under both comparison designs.

\begin{table}[ht]
\centering
\caption{Validation of the Bayesian structural-balance comparison over 100 replicates.}
\label{tab:testing-simulation}
\resizebox{\textwidth}{!}{%
\begin{tabular}{lrrrrrrr}
\hline
Design & $\Pr(B_1>B_2)$ & $\Pr(\Delta_d>0)$ &
$\pi_-$ & $\pi_0$ & $\pi_+$ &
$BF_{01}^{(\epsilon)}$ & Median $\Delta_d$ \\
\hline
Matched NBA--NBA & 0.864 & 0.864 & 0.113 & 0.050 & 0.836 & 0.317 & 3.46 \\
Heterogeneous NBA--NFL & 0.829 & 0.829 & 0.143 & 0.062 & 0.795 & 0.397 & 3.15 \\
\hline
\end{tabular}}
\end{table}

The contrast between the two designs is itself informative. Swapping the NBA
schedule for the much sparser NFL schedule does not flip the inferred
ordering of the two systems, but it does weaken the posterior evidence. The
average probability assigned to the correct practical direction falls from
0.836 to 0.795, while the mass given to the opposite direction and to the
equivalence region grows. Schedule density, in other words, changes how
confident the comparison is, not what it concludes on the common
structural-balance scale.

The interval Bayes factors average 0.317 under the matched design and 0.397
under the heterogeneous one. Both sit below one, so the posterior does not
favor practical equivalence relative to the prior in this directionally
separated experiment, though neither comes close to certainty. That is about
what one should expect from a single season, especially when one system
supplies only 272 comparisons. It also explains why posterior probabilities
serve the purpose better than a flat declaration that one league is or is
not more balanced.

In sum, the second simulation supports the framework's main comparative
claim. Structural balance can be compared across systems with different
numbers of teams, tier configurations, and comparison graphs, because $B$ is
defined independently of all of these. Schedule differences surface instead
through posterior uncertainty. This is precisely the situation the sports
application presents: two leagues can be placed on the same structural scale
even though one of them offers far more information about its latent
competitive structure.

\section{Comparing Competitive Balance and Tier Structure in the NBA and NFL}
\label{sec:application-nba-nfl}

We apply the proposed MFM--BT model to the 2025--2026 National Basketball
Association (NBA) regular season and the 2025 National Football League (NFL)
regular season. Each game enters the analysis as a paired comparison between
the home and away teams, and each league has its own home-advantage parameter
on the common reference scale $\tau_0=1$. NFL postseason games are excluded,
as is the additional Knicks--Spurs special event in the NBA.

For each league, we run two independently initialized MCMC chains. Each chain
contains 4,000 iterations, with the first 1,000 iterations discarded as
burn-in and every third subsequent iteration retained. This yields 1,000
posterior draws for posterior summaries. We
use $\lambda_K=1$ and $\gamma=0.4$ for the MFM prior, and assess convergence
using trace plots, running posterior means, and Gelman--Rubin diagnostics for
$B$ and the home-advantage parameter.

The two schedules differ sharply in density. The NBA season has each of its
30 teams play 82 games, giving 1,230 regular-season comparisons in total.
The NFL, with 32 teams, plays only 17 games apiece, for 272 comparisons.
The NBA therefore provides more than four times as many pairwise observations
on relative team strength, while the NFL season is short enough that a single
game carries considerable weight. This contrast between a long, dense
schedule and a short, sparse one makes the pairing well suited to studying
how schedule information affects what can be learned about competitive
structure.

\subsection{Structural balance and cross-league comparison}

Posterior summaries for the two leagues appear in Table~\ref{tab:nba-nfl-b}.
The posterior mean of $B$ is 0.482 for the NBA and 0.473 for the NFL. The
NBA estimate is nominally higher, but the difference is small relative to
posterior uncertainty: the 90\% credible intervals are
$[0.240,0.681]$ and $[0.205,0.693]$, respectively. The posterior mean of the
home-advantage parameter is 0.274 for the NBA and 0.196 for the NFL. The
corresponding 90\% credible intervals are $[0.172,0.381]$ and
$[-0.030,0.423]$. Thus, the NBA shows a larger estimated home advantage,
although the intervals overlap substantially and the NFL interval includes
zero.

\begin{table}[ht]
\centering
\caption{Posterior summaries of structural balance, home advantage, and
latent tier structure for the NBA and NFL.}
\label{tab:nba-nfl-b}
\begin{tabular}{lrrrrrrr}
\hline
League & $n$ & $M$ & $\mathbb{E}(B\mid Y)$ & 90\% CrI for $B$ &
$\mathbb{E}(h\mid Y)$ & 90\% CrI for $h$ & Dahl $T$ \\
\hline
NBA & 30 & 1,230 & 0.482 & [0.240, 0.681] & 0.274 &
[0.172, 0.381] & 3 \\
NFL & 32 & 272   & 0.473 & [0.205, 0.693] & 0.196 &
[-0.030, 0.423] & 2 \\
\hline
\end{tabular}
\end{table}

To compare the leagues formally, we use the dispersion-scale procedure from
Section~\ref{sec:testing}. Define
\[
\Delta_d
=
d(B_{\mathrm{NFL}})
-
d(B_{\mathrm{NBA}}),
\]
so positive values indicate greater structural balance in the NBA. We
calibrate the practical margin by allowing a 0.25-unit difference in the RMS
tier separation
\[
s(B)=\sqrt{d(B)}
\]
around $B_0=0.5$, which with $\tau_0=1$ gives the conservative
dispersion-scale margin $\epsilon_d=0.77$.

The posterior median of $\Delta_d$ is 0.04, with a 90\% credible interval of
$[-3.91,5.49]$. The posterior probability that the NBA is more structurally
balanced is
\[
\Pr(\Delta_d>0\mid Y)=0.510,
\]
essentially a coin flip. Posterior mass spreads across all three practical
regions,
\[
(\pi_-,\pi_0,\pi_+)
=
(0.315,0.340,0.345),
\]
rather than concentrating on a meaningful difference in either direction, and
the interval Bayes factor $BF_{01}=1.42$ offers only modest evidence for
practical similarity relative to the prior. The data, in other words, support
neither a directional difference in structural balance nor strong practical
equivalence; one season simply does not contain enough information for the
latter.

\begin{table}[ht]
\centering
\caption{Bayesian comparison of NBA and NFL structural balance on the
dispersion scale.}
\label{tab:nba-nfl-test}
\begin{tabular}{lrrrrrrr}
\hline
Median $\Delta_d$ & 90\% CrI & $\Pr(\Delta_d>0)$ &
$\pi_-$ & $\pi_0$ & $\pi_+$ & $BF_{01}$ \\
\hline
0.04 & [-3.91,5.49] & 0.510 & 0.315 & 0.340 & 0.345 & 1.42 \\
\hline
\end{tabular}
\end{table}

The comparison also highlights a useful property of the balance measure. The
NBA supplies more than four times as many regular-season comparisons as the
NFL, yet this does not change what $B$ means, only how well it is estimated.
The NFL interval is noticeably wider at a similar posterior mean, which is
exactly what one expects when less data are available to separate persistent
strength differences from game-level noise. A league should therefore not be
called more or less balanced merely because its schedule happens to produce
a sharper or noisier standings table.

\subsection{Comparison with traditional competitive-balance measures}
\label{sec:traditional-balance}

For context, we also compute conventional competitive-balance measures used
in sports economics from regular-season win percentages, including the
standard deviation, top-to-bottom range, Herfindahl--Hirschman index (HHI),
and Gini coefficient \citep{FortMaxcy2003,utt2002pitfalls,
owen2007measuring,Manasis2013}. Lower values indicate greater
observed balance. These measures summarize realized standings and therefore
address a different question from the structural balance parameter. In
particular, they do not separate persistent differences in underlying team
strength from game-level variation, and they do not quantify uncertainty in
the latent tier structure.

\begin{table}[ht]
\centering
\caption{Traditional competitive-balance measures calculated from
regular-season standings.}
\begin{tabular}{lrrrr}
\hline
League & SD of win percentage & Range & HHI & Gini \\
\hline
NBA & 0.169 & 0.573 & 0.0370 & 0.188 \\
NFL & 0.204 & 0.647 & 0.0363 & 0.230 \\
\hline
\end{tabular}\label{tab:nba-nfl-traditional}
\end{table}

The conventional measures mostly point toward greater observed parity in the
NBA (Table~\ref{tab:nba-nfl-traditional}). The standard deviation of win
percentages is 0.169 in the NBA against 0.204 in the NFL, and the ranges are
0.573 and 0.647. The Gini coefficients, 0.188 and 0.230, tell the same
story. Only the HHI departs slightly, at 0.0370 for the NBA and 0.0363 for
the NFL. The conventional toolkit thus does not even agree with itself about
the ranking, though most of its members see more dispersion in the NFL
standings.

The model-based analysis cuts across this. Despite the wider spread of
observed NFL win percentages under most conventional measures, the posterior
distributions of $B$ give little reason to think the leagues differ in
underlying structural balance. Some of what the standings show as an NFL
disadvantage is plausibly the product of a much shorter schedule and the
game-level variation that comes with it, rather than genuinely larger
persistent heterogeneity. Observed parity and latent structural balance,
it seems, need not tell the same story.

\subsection{Latent competitive tiers and team rankings}

The structural balance parameter measures overall heterogeneity, but it does
not describe how teams are organized within a league. We therefore also
examine the estimated competitive tiers. Because mixture labels are arbitrary, we order the tiers according to the
Dahl-selected tier abilities, with Tier 1 denoting the strongest competitive
tier. The Dahl representative partition identifies
three tiers in the NBA with 6, 14, and 10 teams, respectively, and two tiers
in the NFL with 16 teams each
(Table~\ref{tab:nba-nfl-tiers}).

\begin{table}[ht]
\centering
\caption{MFM--BT Dahl representative tier composition for the NBA and NFL.
Tier 1 denotes the strongest competitive tier.}

\begin{tabularx}{\textwidth}{lX}
\hline
League and tier & Teams \\
\hline
NBA Tier 1 (higher) &
Boston Celtics; Denver Nuggets; Detroit Pistons; New York Knicks;
Oklahoma City Thunder; San Antonio Spurs. \\

NBA Tier 2 (middle) &
Atlanta Hawks; Charlotte Hornets; Cleveland Cavaliers;
Golden State Warriors; Houston Rockets; LA Clippers;
Los Angeles Lakers; Miami Heat; Minnesota Timberwolves;
Orlando Magic; Philadelphia 76ers; Phoenix Suns;
Portland Trail Blazers; Toronto Raptors. \\

NBA Tier 3 (lower) &
Brooklyn Nets; Chicago Bulls; Dallas Mavericks; Indiana Pacers;
Memphis Grizzlies; Milwaukee Bucks; New Orleans Pelicans;
Sacramento Kings; Utah Jazz; Washington Wizards. \\

NFL Tier 1 (higher) &
Buffalo Bills; Chicago Bears; Denver Broncos; Detroit Lions;
Green Bay Packers; Houston Texans; Indianapolis Colts;
Jacksonville Jaguars; Los Angeles Chargers; Los Angeles Rams;
Minnesota Vikings; New England Patriots; Philadelphia Eagles;
Pittsburgh Steelers; San Francisco 49ers; Seattle Seahawks. \\

NFL Tier 2 (lower) &
Arizona Cardinals; Atlanta Falcons; Baltimore Ravens; Carolina Panthers;
Cincinnati Bengals; Cleveland Browns; Dallas Cowboys; Kansas City Chiefs;
Las Vegas Raiders; Miami Dolphins; New Orleans Saints; New York Giants;
New York Jets; Tampa Bay Buccaneers; Tennessee Titans;
Washington Commanders. \\
\hline
\end{tabularx}
\label{tab:nba-nfl-tiers}
\end{table}

The NBA shows a more differentiated competitive structure. Six teams make up
Tier 1, a large middle group of 14 sits in Tier 2, and 10 teams fall to
Tier 3. Much of the league, in other words, is packed into the middle, with
a relatively small group pulling away at the top. This is also where the
tier structure and the global balance parameter part company: a league can
be highly balanced overall and still contain a recognizable group of leading
teams. The tier compositions also say things the raw standings do not. Tier 1 puts
the defending champion Oklahoma City Thunder and the usual contenders,
Boston and Denver, alongside two teams a standings-only reading would
overlook. Detroit spent much of the season playing better than its win total
suggested, and San Antonio belonged to the elite group on latent strength
well before its record caught up. The Knicks sit in that group too, which
shows the model looking past short streaks to how teams actually compare.
Working the other way, a few famous names, the Lakers and Warriors among
them, land in the broad middle tier rather than the elite one, and a
win-percentage table flattens that distinction into one long ranking. The
practical reading for a front office is simple: about half the league is a
middle band where games are close to a coin flip, and a six-team group
forms the realistic championship tier. That distinction shapes how a season
should be judged, and no standings table makes it visible.

The NFL partition is coarser, splitting the league evenly into two tiers of
16. That should not be read as a simpler competitive structure. With only
272 regular-season games against the NBA's 1,230, the data simply carry far
less information about teams of similar ability. The broad NFL tiers reflect
both the organization the data support and the heavier uncertainty that a
sparse schedule brings.
The NFL tiers carry the same kind of information. Tier 1 contains the
eventual Super Bowl champion Seattle and the AFC champion New England, along
with teams whose seasons ended earlier than their strength deserved, Buffalo
and the Chargers being the clearest cases. It also holds the season's
surprise division winners, including Chicago, whose arrival in the top group
the model caught even though the franchise had spent years near the bottom.
Tier 2, meanwhile, collects teams whose reputations run ahead of their
present strength. Kansas City and Baltimore both fall below the midpoint on
latent ability despite championship pedigrees that say otherwise. In a
seventeen-game season, a couple of close losses or one injury at the wrong
time can push a team like that out of January, and the tiers capture the
fact that they never separated from the league's middle. The honest summary
of an NFL season, it turns out, is a large top group of broadly comparable
clubs, a handful of genuine stragglers, and a lot of doubt in between, not
the clean thirty-two-step ordering the standings imply.

The comparison also shows why the number of tiers should not be used as a
measure of overall competitive balance. The NBA has three estimated tiers and
the NFL has two, while their posterior estimates of $B$ are very similar.
The tier structure describes how teams are organized into competitive levels,
whereas $B$ measures the magnitude of heterogeneity among those levels.
Therefore, more tiers do not necessarily imply less competitive balance.

\begin{figure}[ht]
\centering
\includegraphics[width=0.98\textwidth]{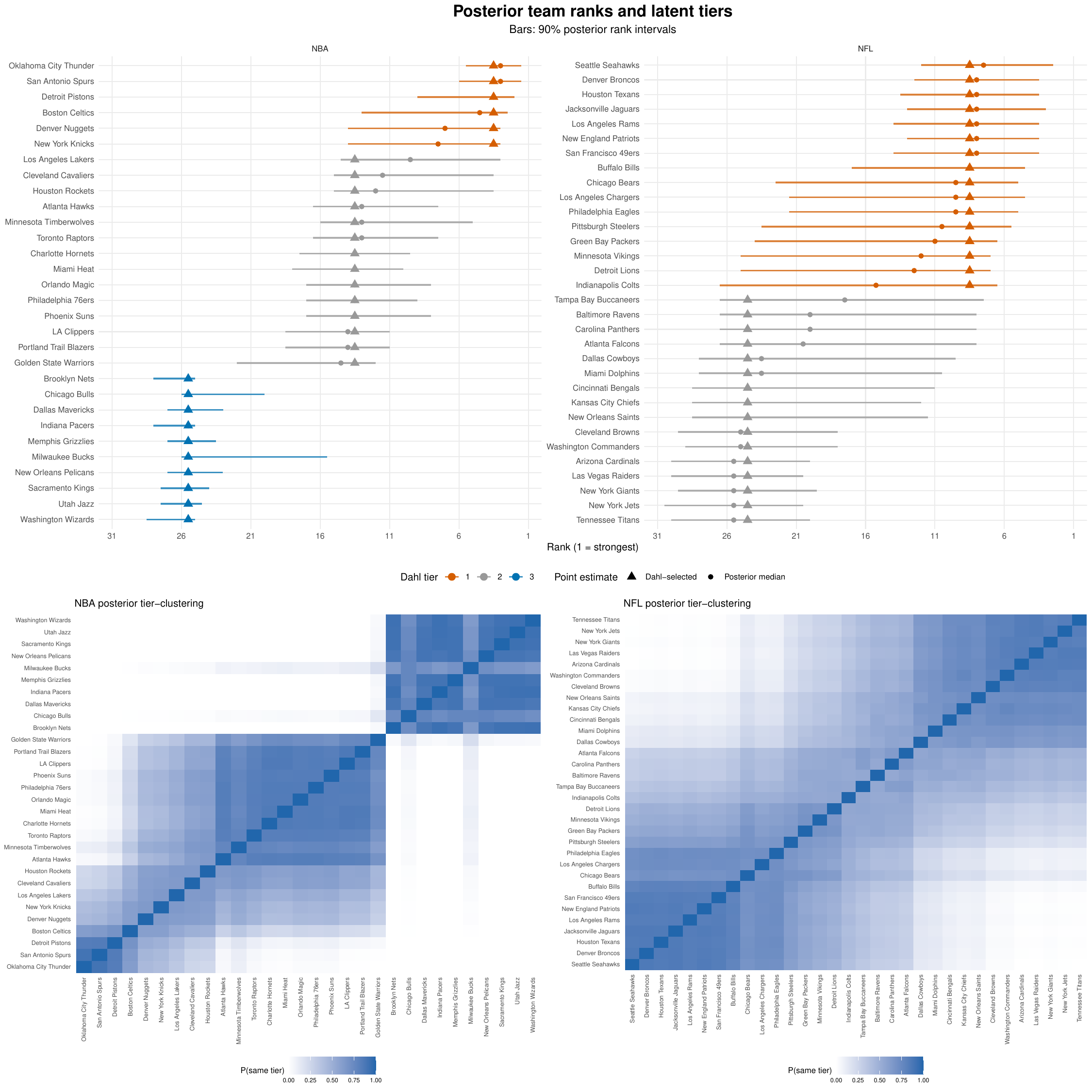}
\caption{Posterior team rankings, Dahl representative tiers, and posterior
mean co-clustering matrices for the NBA and NFL. Circles denote posterior
median ranks, triangles denote ranks from the Dahl representative partition,
and horizontal bars represent 90\% posterior credible intervals for team
ranks. Team colors indicate the Dahl representative tiers, with Tier 1
denoting the strongest competitive tier. The heatmaps display the posterior
mean co-clustering matrices, where each entry is the posterior mean of the
indicator that two teams belong to the same latent tier. Rank 1 denotes the
strongest team.}
\label{fig:nba-nfl-ranks}
\end{figure}

Figure~\ref{fig:nba-nfl-ranks} provides a more detailed view of the posterior
ranking and clustering results. The rank intervals describe uncertainty in
the relative positions of individual teams, while the posterior mean
co-clustering matrices summarize uncertainty in whether pairs of teams belong
to the same latent competitive tier.

The posterior mean co-clustering matrices provide information that is not
available from a single representative partition. Teams near a tier boundary
may have lower posterior co-clustering probabilities with teams in their
assigned tier and nonnegligible probabilities of clustering with teams in a
neighboring tier. The Dahl tiers should therefore be interpreted as
representative summaries of posterior clustering uncertainty rather than as
fixed classifications of team quality.

\subsection{Agreement with observed standings and practical implications}

As a descriptive check, we compare posterior median ranks with observed
regular-season ranks based on winning percentage, using point differential
as a transparent tie-breaker. Since both rankings come from the same games,
this measures agreement with the observed season rather than out-of-sample
prediction.

\begin{table}[ht]
\centering
\caption{Agreement between MFM--BT posterior ranks and observed
regular-season ranks.}
\label{tab:nba-nfl-rank-correlation}
\begin{tabular}{lrrrrr}
\hline
League & Spearman $\rho$ & Kendall $\tau$ & Pearson $r$ &
Mean absolute rank error & Top-five overlap \\
\hline
NBA & 0.978 & 0.924 & 0.949 & 2.35 & 5/5 \\
NFL & 0.990 & 0.944 & 0.959 & 2.38 & 4/5 \\
\hline
\end{tabular}
\end{table}

The model ranks teams much the way the standings do. The Spearman correlation
is 0.978 for the NBA and 0.990 for the NFL, and the Kendall correlations are
0.924 and 0.944. The model's top five also matches the observed NBA top five
exactly, and four of the observed NFL top five. So the model keeps the
ordering we see in the season results, but adds something the standings
cannot: uncertainty intervals and information about which teams belong to the
same competitive group.

The results leave us with a few observations. First, the usual measures
cannot even settle which league is more balanced. Most of the dispersion
measures point to the NBA, the HHI points slightly to the NFL, and our model
finds no meaningful difference in structural balance. Much depends on what
one chooses to look at: the results on the court and field, or the strength
of the teams underneath them.
It also becomes clear that equal balance does not mean the same kind of
competition. The NBA breaks into three tiers, with most clubs crowded into a
broad middle and a small elite group above them, while the NFL settles into
two wide groups. A policy aimed at parity might close the gaps between teams
without changing how they bunch together, or it might reshuffle the groups
without narrowing the gaps, and it is worth knowing which of the two a given
policy actually does.
Finally, the schedule puts a ceiling on what any of this can tell us. An NBA
season simply contains far more information about how teams compare, whereas
an NFL season leaves real doubt about teams of similar strength and about
where one tier gives way to the next. A comparison built only on standings
risks mistaking that doubt for a genuine difference in balance.
On the whole, the NBA and NFL look much alike in global structural balance
but quite different in how their competition is organized, and in how
confidently that organization can be described. Balance, tier structure, and
schedule information are separate features of a league. Treating them as
such makes for better comparisons than leaning on standings, conventional
indices, or tier counts alone.

\section{Discussion and Conclusion}
\label{sec:conclusion}

We have developed a Bayesian paired-comparison framework for studying competitive balance that separates global structural heterogeneity from latent competitive tier structure. The MFM prior learns the number and membership of occupied tiers, while the balance parameter $B$ quantifies how strongly those tiers are separated. Our theoretical results give $B$  a direct interpretation in terms of between-tier dispersion and show how overall team-level heterogeneity depends jointly on the size of tier separation and the prevalence of cross-tier comparisons. Because the framework separates the underlying competitive structure from the information contained in the observed schedule, balance can be compared across systems with different numbers of teams and games.
The simulations show that the method recovers latent competitive structure accurately when the comparison design is informative enough, and that estimates of structural balance remain comparatively stable even as schedules thin out. The NBA and NFL application reinforces the distinction between global balance and competitive stratification. Conventional standings-based measures generally suggest greater observed parity in the NBA, yet the posterior analysis finds little evidence of a meaningful difference in underlying structural balance. The two leagues are organized differently all the same: the NBA splits into three latent competitive tiers, while the NFL falls into two broader ones, and the greater uncertainty in the NFL reflects its much sparser schedule.

Several extensions are worth exploring. A longitudinal version of the model could let $B$, team abilities, and tier memberships drift across seasons, which would make it possible to characterize changes in parity, movement between competitive tiers, and the effects of league policies on competitive structure. On the theory side, it would be useful to understand how the topology of the comparison graph shapes estimation. Beyond plain connectivity, properties such as graph density, degree heterogeneity, spectral connectivity, and bottlenecks may govern how precisely abilities, tiers, and structural balance can be learned, and could inform the design of schedules that carry more information.
One further direction is to relax the transitive structure of the Bradley--Terry model. Sports competitions often show systematic matchup patterns: team A tends to beat team B, team B tends to beat team C, but team C tends to beat team A. A one-dimensional latent ability cannot represent such cycles. A natural next step is to augment the global ability component with structured matchup effects that capture nontransitive patterns while keeping measures of overall strength and balance interpretable. That would give a richer account of both hierarchical and cyclic structure in competitive systems.

\bibliographystyle{apalike}
\bibliography{ref}

@article{miller2018mixture,
  author  = {Miller, Jeffrey W. and Harrison, Matthew T.},
  title   = {Mixture Models with a Prior on the Number of Components},
  journal = {Journal of the American Statistical Association},
  year    = {2018},
  volume  = {113},
  number  = {521},
  pages   = {340--356},
  doi     = {10.1080/01621459.2016.1255636}
}

@article{polson2013bayesian,
  author  = {Polson, Nicholas G. and Scott, James G. and Windle, Jesse},
  title   = {Bayesian Inference for Logistic Models Using P{\'o}lya--Gamma Latent Variables},
  journal = {Journal of the American Statistical Association},
  year    = {2013},
  volume  = {108},
  number  = {504},
  pages   = {1339--1349},
  doi     = {10.1080/01621459.2013.829001}
}

@article{bradley1952rank,
  author  = {Bradley, Ralph Allan and Terry, Milton E.},
  title   = {Rank Analysis of Incomplete Block Designs: {I}. The Method of Paired Comparisons},
  journal = {Biometrika},
  year    = {1952},
  volume  = {39},
  number  = {3/4},
  pages   = {324--345},
  doi     = {10.1093/biomet/39.3-4.324}
}

@article{basini2023assessing,
  author  = {Basini, Francesca and Tsouli, Vasiliki and Ntzoufras, Ioannis and Friel, Nial},
  title   = {Assessing Competitive Balance in the {E}nglish {P}remier {L}eague for over Forty Seasons Using a Stochastic Block Model},
  journal = {Journal of the Royal Statistical Society Series A: Statistics in Society},
  year    = {2023},
  volume  = {186},
  number  = {3},
  pages   = {530--556},
  doi     = {10.1093/jrsssa/qnad007}
}

@article{neal2000markov,
  author  = {Neal, Radford M.},
  title   = {Markov Chain Sampling Methods for {D}irichlet Process Mixture Models},
  journal = {Journal of Computational and Graphical Statistics},
  year    = {2000},
  volume  = {9},
  number  = {2},
  pages   = {249--265},
  doi     = {10.1080/10618600.2000.10474879}
}

@article{owen2007measuring,
  title   = {Measuring Competitive Balance in Professional Team Sports
             Using the Herfindahl--Hirschman Index},
  author  = {Owen, P. Dorian and Ryan, Michael and Weatherston, Clayton R.},
  journal = {Review of Industrial Organization},
  volume  = {31},
  number  = {4},
  pages   = {289--302},
  year    = {2007},
  doi     = {10.1007/s11151-008-9157-0}
}

@article{dong2026statistical,
  title   = {Statistical Ranking with Dynamic Covariates},
  author  = {Dong, Pinjun and Han, Ruijian and Jiang, Binyan and Xu, Yiming},
  journal = {Journal of the Royal Statistical Society Series B: Statistical Methodology},
  volume  = {88},
  number  = {1},
  pages   = {221--238},
  year    = {2026},
  doi     = {10.1093/jrsssb/qkaf048}
}

@incollection{dahl2006model,
  title     = {Model-Based Clustering for Expression Data via a Dirichlet Process Mixture Model},
  author    = {Dahl, David B.},
  booktitle = {Bayesian Inference for Gene Expression and Proteomics},
  editor    = {Do, Kim-Anh and M{\"u}ller, Peter and Vannucci, Marina},
  pages     = {201--218},
  publisher = {Cambridge University Press},
  year      = {2006},
  doi       = {10.1017/CBO9780511584589.011}
}

@article{masarotto2012ranking,
  title={The ranking lasso and its application to sport tournaments},
  author={Masarotto, Guido and Varin, Cristiano},
  journal={The Annals of Applied Statistics},
  volume={6},
  number={4},
  pages={1949--1970},
  year={2012},
  publisher={Institute of Mathematical Statistics},
  doi={10.1214/12-AOAS581}
}

@article{pearce2025bayesian,
  author  = {Pearce, Michael and Erosheva, Elena A.},
  title   = {Bayesian Rank-Clustering},
  journal = {Psychometrika},
  volume  = {90},
  number  = {3},
  pages   = {904--931},
  year    = {2025},
  doi     = {10.1017/psy.2025.10014}
}

@book{Luce1959,
  title={Individual Choice Behavior: A Theoretical Analysis},
  author={Luce, R Duncan},
  year={1959},
  publisher={Wiley}
}

@article{glickman1999parameter,
  title={Parameter estimation in large dynamic paired comparison experiments},
  author={Glickman, Mark E},
  journal={Journal of the Royal Statistical Society Series C: Applied Statistics},
  volume={48},
  number={3},
  pages={377--394},
  year={1999},
  publisher={Oxford University Press}
}

@article{mollica2021bayesian,
  title={Bayesian analysis of ranking data with the Extended Plackett--Luce model},
  author={Mollica, Cristina and Tardella, Luca},
  journal={Statistical Methods \& Applications},
  volume={30},
  number={1},
  pages={175--194},
  year={2021},
  publisher={Springer}
}

@article{hunter2004mm,
  title={{MM} algorithms for generalized {B}radley-{T}erry models},
  author={Hunter, David R},
  journal={The Annals of Statistics},
  volume={32},
  number={1},
  pages={384--406},
  year={2004},
  publisher={Institute of Mathematical Statistics}
}

@article{davidson1970extending,
  title={On extending the Bradley-Terry model to accommodate ties in paired comparison experiments},
  author={Davidson, Roger R},
  journal={Journal of the American Statistical Association},
  volume={65},
  number={329},
  pages={317--328},
  year={1970},
  publisher={Taylor \& Francis}
}

@article{dunson2008bayesian,
  title={Bayesian nonparametric inference on stochastic ordering},
  author={Dunson, David B and Peddada, Shyamal D},
  journal={Biometrika},
  volume={95},
  number={4},
  pages={859--874},
  year={2008},
  publisher={Oxford University Press}
}

@article{utt2002pitfalls,
  title={Pitfalls to measuring competitive balance with Gini coefficients},
  author={Utt, Joshua and Fort, Rodney},
  journal={Journal of Sports Economics},
  volume={3},
  number={4},
  pages={367--373},
  year={2002},
  publisher={Sage Publications Sage CA: Thousand Oaks, CA}
}

@article{li2022bayesian,
  title={Bayesian analysis of rank data with covariates and heterogeneous rankers},
  author={Li, Xinran and Yi, Dingdong and Liu, Jun S},
  journal={Statistical Science},
  volume={37},
  number={1},
  pages={1--23},
  year={2022},
  publisher={Institute of Mathematical Statistics}
}

@inproceedings{busse2007cluster,
  title={Cluster analysis of heterogeneous rank data},
  author={Busse, Ludwig M and Orbanz, Peter and Buhmann, Joachim M},
  booktitle={Proceedings of the 24th international conference on Machine learning},
  pages={113--120},
  year={2007}
}

@article{murphy2003mixtures,
  title={Mixtures of distance-based models for ranking data},
  author={Murphy, Thomas Brendan and Martin, Donal},
  journal={Computational statistics \& data analysis},
  volume={41},
  number={3-4},
  pages={645--655},
  year={2003},
  publisher={Elsevier}
}

@article{deng2014bayesian,
  title={Bayesian aggregation of order-based rank data},
  author={Deng, Ke and Han, Simeng and Li, Kate J and Liu, Jun S},
  journal={Journal of the American Statistical Association},
  volume={109},
  number={507},
  pages={1023--1039},
  year={2014},
  publisher={Taylor \& Francis}
}

@book{marden1996analyzing,
  title={Analyzing and modeling rank data},
  author={Marden, John I},
  year={1996},
  publisher={CRC Press}
}

@article{FortQuirk1995,
  author  = {Fort, Rodney and Quirk, James},
  title   = {Cross-Subsidization, Incentives, and Outcomes in Professional Team Sports Leagues},
  journal = {Journal of Economic Literature},
  volume  = {33},
  number  = {3},
  pages   = {1265--1299},
  year    = {1995},
  month   = sep,
  url     = {https://www.jstor.org/stable/2729122}
}

@article{Vrooman1995,
  title={A general theory of professional sports leagues},
  author={Vrooman, John},
  journal={Southern Economic Journal},
  volume={61},
  number={4},
  pages={971--990},
  year={1995}
}

@article{Szymanski2001,
  title={Income inequality, competitive balance and the attractiveness of team sports},
  author={Szymanski, Stefan},
  journal={Economic Journal},
  volume={111},
  number={469},
  pages={69--84},
  year={2001}
}

@article{Szymanski2003,
  title={The economic design of sporting contests},
  author={Szymanski, Stefan},
  journal={Journal of Economic Literature},
  volume={41},
  number={4},
  pages={1137--1187},
  year={2003}
}

@article{Zimbalist2002,
  title={Competitive balance in sports leagues: An introduction},
  author={Zimbalist, Andrew},
  journal={Journal of Sports Economics},
  volume={3},
  number={2},
  pages={111--121},
  year={2002}
}

@article{FortMaxcy2003,
  title={Competitive balance in sports leagues: An introduction},
  author={Fort, Rodney and Maxcy, Joel},
  journal={Journal of Sports Economics},
  volume={4},
  number={2},
  pages={154--160},
  year={2003}
}

@article{Manasis2013,
  author  = {Manasis, Vasilis and Avgerinou, Vasiliki and Ntzoufras, Ioannis and Reade, J. James},
  title   = {Quantification of Competitive Balance in {E}uropean Football: Development of Specially Designed Indices},
  journal = {IMA Journal of Management Mathematics},
  volume  = {24},
  number  = {3},
  pages   = {363--375},
  year    = {2013},
  doi     = {10.1093/imaman/dps014}
}

@article{ManasisNtzoufras2014,
  author  = {Manasis, Vasilis and Ntzoufras, Ioannis},
  title   = {Between-Seasons Competitive Balance in {E}uropean Football: Review of Existing and Development of Specially Designed Indices},
  journal = {Journal of Quantitative Analysis in Sports},
  volume  = {10},
  number  = {2},
  pages   = {139--152},
  year    = {2014},
  doi     = {10.1515/jqas-2013-0107}
}

@article{Manasis2022,
  title={Competitive balance measures and the uncertainty of outcome hypothesis in European football},
  author={Manasis, Vasilis and Ntzoufras, Ioannis and Reade, J. James},
  journal={IMA Journal of Management Mathematics},
  volume={33},
  number={1},
  pages={19--52},
  year={2022}
}

@article{tian2024spectral,
  title={A spectral approach for the dynamic Bradley--Terry model},
  author={Tian, Xinyu and Shi, Jian and Shen, Xiaotong and Song, Kai},
  journal={Stat},
  volume={13},
  number={3},
  pages={e722},
  year={2024},
  publisher={Wiley Online Library}
}

@article{caron2012efficient,
  title={Efficient Bayesian inference for generalized Bradley--Terry models},
  author={Caron, Francois and Doucet, Arnaud},
  journal={Journal of Computational and Graphical Statistics},
  volume={21},
  number={1},
  pages={174--196},
  year={2012},
  publisher={Taylor \& Francis}
}

@article{pearce2025modeling,
  title   = {Modeling Preferences: A Bayesian Mixture of Finite Mixtures for Rankings and Ratings},
  author  = {Pearce, Michael and Erosheva, Elena A.},
  journal = {Journal of the American Statistical Association},
  volume  = {120},
  number  = {551},
  pages   = {1621--1632},
  year    = {2025},
  doi     = {10.1080/01621459.2024.2444700}
}

@article{santi2026bradley,
  title   = {The Bradley--Terry Stochastic Block Model},
  author  = {Santi, Lapo and Friel, Nial},
  journal = {The Annals of Applied Statistics},
  volume  = {20},
  number  = {2},
  pages   = {1766--1787},
  year    = {2026},
  doi     = {10.1214/26-AOAS2193}
}

@inproceedings{kulis2012revisiting,
  author    = {Kulis, Brian and Jordan, Michael I.},
  title     = {Revisiting {K}-Means: New Algorithms via Bayesian Nonparametrics},
  booktitle = {Proceedings of the 29th International Conference on Machine Learning},
  pages     = {1131--1138},
  year      = {2012}
}

@book{mclachlan2000finite,
  title     = {Finite Mixture Models},
  author    = {McLachlan, Geoffrey J. and Peel, David},
  year      = {2000},
  publisher = {John Wiley \& Sons},
  address   = {New York},
  series    = {Wiley Series in Probability and Statistics},
  doi       = {10.1002/0471721182}
}

@article{hubert1985comparing,
  title   = {Comparing Partitions},
  author  = {Hubert, Lawrence and Arabie, Phipps},
  journal = {Journal of Classification},
  volume  = {2},
  number  = {1},
  pages   = {193--218},
  year    = {1985},
  doi     = {10.1007/BF01908075}
}

\end{document}